\documentclass[prl, twocolumn,  amsmath, showpacs]{revtex4}

\usepackage{graphicx}

\newcommand{\HGeo}{{}_2\!F_1}
\begin{document}
 \title{Separable sequences in Bianchi I loop quantum cosmology}

\author{Daniel Cartin}
\email{cartin@naps.edu}\affiliation{Naval Academy Preparatory School, 197 Elliot Street, Newport, Rhode Island 02841}

\author{Gaurav Khanna}
\email{gkhanna@UMassD.Edu}
\affiliation{Physics Department, University of Massachusetts at Dartmouth, North Dartmouth, Massachusetts 02747}

\date{\today}
\begin{abstract}
In this paper, we discuss the properties of one-parameter sequences that arise when solving the Hamiltonian constraint in Bianchi I loop quantum cosmology using a separation of variables method. In particular, we focus on finding an expression for the sequence for all real values of the parameter, and discuss the pre-classicality of this function. We find that the behavior of these  preclassical sequences imply time asymmetry on either side of the classical singularity in Bianchi I cosmology.  

\end{abstract}


\maketitle

Loop quantum cosmology (LQC)~\cite{lqc} has been developed as a symmetry reduced model of quantum geometry~\cite{rov}, to test the viability of the latter as a theory of quantum gravity. Analogous to minisuperspace models, LQC reduces the infinite degrees of freedom in the full theory down to a finite, quantum mechanical theory. This allows study of various models of cosmological interest, such as isotropic Friedman-Robertson-Walker (FRW) space-times. In particular, this has led to the emergence of a quantum gravity mechanism for inflation~\cite{inf}, evolution through the classical singularity~\cite{boj02} or oscillatory universes~\cite{osc}, as well as the development of effective semi-classical Hamiltonians~\cite{ban-dat05}. Developments in the toy model of LQC can also help direct inquiry in the full model~\cite{bru-thi05}.

The quantum Hamiltonian constraint in LQC is a difference equation, relating discrete eigenstates of the triad operators. The generic solution to a difference equation, however, has rather unphysical properties; in particular, as the triad parameter is increased by discrete steps, the coefficients of the wave function expansion will alternate in sign as they increase in magnitude. This can be interpreted physically as quantum effects becoming important even in large-sized universes. At the present time, there is no physical inner product in LQC to eliminate unphysical states with large quantum effects. This has led to the notion of {\it pre-classicality}~\cite{boj01}, where states with oscillatory behavior far away from the singularity are eliminated from consideration. Generating function techniques~\cite{car-kha-boj04, car-kha05}, useful in solving the multi-parameter recursion relations arising in the anisotropic models, can implement pre-classicality by placing boundary conditions on the function associated to the desired sequence.

However, these methods have been used to find the solutions only for integer values of the parameters (although there exists a general solution in the isotropic case for all real values~\cite{boj02}; see also~\cite{nou-per-van05}). This raises the question of the existence of solutions for non-integer values in anisotropic models. Recent research has addressed this by treating each non-integer sequence distinctly~\cite{dat05}, rather than as part of a general solution for the recursion relation. In this work, we will find that obtaining the generating function allows us to find a {\it single} solution applicable for {\it all} values of the parameters. We do this by separating the anisotropic recursion relations into products of one-parameter equations. Models where this is possible include the Bianchi I system~\cite{car-kha05} and its locally rotationally symmetric (LRS) version~\cite{car-kha-boj04}; in addition, the Kantowski-Sachs model, relevant for the interior of the Schwarzschild black hole, is a similar case~\cite{boj-com}. Because the same one-parameter recursion relation appears in these contexts, it allows us to study the general properties of wave functions as they pass from one side of the singularity to the other. Then we can examine their pre-classicality over the range of all triad eigenvalues, as well as differences in their behavior for positive and negative parameters.

In particular, we look at the Bianchi I model, the anisotropic generalization of the spatially flat FRW cosmology; a complete explanation of its quantization in the LQC context appears in~\cite{boj03}. As mentioned before, the quantized Hamiltonian constraint, acting on eigenstates of the triad operators, gives a difference equation relating the coefficients $s(m_1, m_2, m_3)$ of the wave function in the triad basis, $\psi = \sum s(m_1, m_2, m_3) |m_1, m_2, m_3 \rangle$. Note that the parameters $m_k$ take real values, not just integers. Because of the freedom to change the sign of two tetrad vectors, the coefficients $s(m_1, m_2, m_3)$ in the full Bianchi I model must satisfy
\begin{align*}
s(m_1, m_2, m_3) &= s(-m_1, -m_2, m_3) = s(-m_1, m_2, -m_3) \\
&= s(m_1, -m_2, -m_3).
\end{align*}
There is a similar condition for the LRS case and its coefficients $s(m, n)$, with $m$ is the parameter in the direction of the rotational symmetry. This remaining gauge symmetry allows us to let one of the parameters run from $-\infty$ to $\infty$, with the others restricted to non-negative values. Here, we choose the classical singularity for Bianchi I to be $m_3 = 0$, and $n=0$ is used for Bianchi I LRS.

When the Hamiltonian constraint is quantized, we find that the coefficients $s(m_1, m_2, 0)$ representing the zero volume classical singularity fall out of the recursion relation. This is due to the difference equation containing functions of the volume $V(m_1, m_2, m_3) = (\frac{1}{2} \gamma \ell _P ^2)^{3/2} \sqrt{| m_1 m_2 m_3 |}$, where $\gamma$ is the Barbero-Immirzi parameter, and $\ell_P$ the Planck length. The actual difference equation is easier to deal with when we make the substitution $t(m_1, m_2, m_3) = V(2m_1, 2m_2, 2m_3) s(2m_1, 2m_2, 2m_3)$, which gives rise to the conditions $t(0, m_2, m_3) = 0$, etc. If we define the difference operator $\delta_1$ as
\[
\delta_1 t (m_1, m_2, m_3) \equiv t (m_1 + 1, m_2, m_3) - t (m_1 - 1, m_2, m_3),
\]
and similarly for $\delta_2$ and $\delta_3$, the Hamiltonian constraint in the full Bianchi I model is given by the difference equation
\begin{equation}
\label{full-eqn}
[ d(m_1) \delta_2 \delta_3 + d(m_2) \delta_1 \delta_3 + d(m_3) \delta_1 \delta_2] t (m_1, m_2, m_3) = 0,
\end{equation}
where $d(n) = \sqrt{|1 + \frac{1}{2n}|} - \sqrt{|1 - \frac{1}{2n}|}$. The factors of $d(m_k)$ arise due to the various functions of the volume $V$ existing in the constraint equation. In the LRS case, the Hamiltonian constraint becomes
\[
[d(n) \delta_m \delta_m + 2 d_2 (m) \delta_m \delta_n] {\tilde t} (m, n) = 0,
\]
with $\delta_m$ and $\delta_n$ are the difference operators for the corresponding parameters, and again there are the conditions ${\tilde t}(0, n) = {\tilde t} (m, 0) = 0$. When we write the equation in terms of a derived sequence $t (m, n) = \delta_m {\tilde t} (m, n)$, then the relation for Bianchi I LRS becomes
\begin{equation}
\label{LRS-eqn}
[d(n) \delta_m + 2 d_2(m) \delta_n] t (m, n) = 0,
\end{equation}
with the function $d_2(m)$ defined as $1/m$ when $m \ne 0$, and $d_2(0) = 0$. Because of the switch from ${\tilde t} (m, n)$ to $t (m, n)$ in the LRS case, $t (0, n)$ is no longer required to be zero.

One method of solving the constraint relations (\ref{full-eqn}) and (\ref{LRS-eqn}) is using a separation of variables technique. In particular, if we look at the LRS case, we can assume the total sequence $t (m, n)$ is the product $a_\lambda (m) b_\lambda (n)$ of two one-dimensional sequences, with a separation parameter $\lambda$. If we consider $n$, the parameter in the direction of the classical singularity for Bianchi I LRS, as an "internal time", then the sequence $b_\lambda (n)$ describes the temporal evolution of the wave function. To solve the full recursion relation, the two sequences must satisfy
\begin{subequations}
\begin{eqnarray}
\label{a-eqn}
\delta_m a_\lambda (m) &=& \frac{2\lambda}{m} a_\lambda (m), \\
\label{b-eqn}
\delta_n b_\lambda (n) &=& - \lambda d(n) b_\lambda (n),
\end{eqnarray}
\end{subequations}
when $m, n \ne 0$ (when $m$ or $n$ equal zero, the right hand sides of the equations are identically zero).  The full Bianchi I case is similar -- assuming the wave function coefficients $t (m_1, m_2, m_3)$ are the product of three one-parameter sequences $c_k (m_k)$, then each must solve
\begin{equation}
\label{c-eqn}
\delta_k c_{\lambda_k} (m_k) = \lambda_k d(m_k) c_{\lambda_k} (m_k),
\end{equation}
with $m_k \ne 0$ and the separation parameters $\lambda_k$ restricted by the relation $\lambda_1 \lambda _2 + \lambda_1 \lambda _3 + \lambda_2 \lambda _3 = 0$. Again, choosing the singularity at $m_3 = 0$ means that the sequence $c_{\lambda_3} (m_3)$ gives the evolution of the wave function through the singularity.

In the regime where $n$ is small, the function $d(n) \simeq (1/2) d_2 (n)$; far away from the singularity, the sequences solving the recursion relations (\ref{a-eqn}), (\ref{b-eqn}) and (\ref{c-eqn}) are essentially the same (up to a scaling of the separation constant $\lambda$). This means that finding the solution for a particular recursion relation is equivalent to finding one for all, at least for large values of the parameters. Closer to where $m_3$ or $n$ are zero, the difference between the functions $d(n)$ and $d_2(m)$ appearing in the recursion relations will be more noticeable. Because the non-polynomial $d(n)$ is more difficult to deal with, we shall first obtain the solution for the difference equation with $d_2(m)$, which has the same asymptotic values as one with $d(n)$, then numerically evolve the sequence through the singularity using the function $d(n)$. Another important point to note is the relative sign between the sequences. For example, a solution for $a_\lambda (m)$ with a value $\lambda$ maps over to a solution to $b_\lambda (m)$ with a separation constant of $-\lambda$. Similarly, the condition  $\lambda_1 \lambda _2 + \lambda_1 \lambda _3 + \lambda_2 \lambda _3 = 0$ for the full Bianchi I constraint means that at least one $\lambda_k$ must have a different sign from the others.

We focus on the recursion relation (\ref{a-eqn}) for the sequence $a_\lambda (m)$ as a specific case, keeping in mind how to generalize this to the other recursion relations as discussed above. To solve this relation, we convert the problem of finding an appropriate sequence to that of finding the solution to a differential equation. This is a method that has been explored in detail elsewhere in the context of LQC~\cite{car-kha-boj04}, so we only provide a summary here. One attempts to find a generating function $F_\lambda (x)$, with a formal power series $F_\lambda (x) = \sum_{m=0} ^\infty a_\lambda (m) x^m$, where the coefficients $a_\lambda (m)$ solve the recursion relation (\ref{a-eqn}). Instead of working directly with the generating function $F_\lambda (x)$ associated with the sequence, it turns out it is simpler to use the function $G_\lambda (x)$, where $F_\lambda (x) = a_\lambda (0) + x G_\lambda (x)$; this auxiliary function must satisfy
\begin{equation}
\label{gen-ODE}
\frac{d}{dx} \biggl[ (1 - x^2) G_\lambda (x) \biggr] - 2\lambda G_\lambda (x) = a_\lambda (0),
\end{equation}
with $a_\lambda(0) = F_\lambda (0)$ the value of the sequence at $m=0$, and $G_\lambda (0) = a_\lambda (1)$. We will see that the values of $a_\lambda (0)$ and $\lambda$ are related for pre-classical sequences. If $\lambda > 0$, then any value of $a_\lambda (0)$ will give pre-classicality; when $\lambda < 0$, then sequences will be smooth on only one side of the singularity when $a_\lambda (0) \ne 0$ and nowhere at all when $a_\lambda (0) = 0$ (it is easily seen that the case $\lambda = 0$ corresponds to a constant sequence).

The situation $a_\lambda (0) = 0$ enforces the boundary conditions at the singularity; the zero-volume part of the wave function drops out of the Hamiltonian constraint at this location. Solving the differential equation (\ref{gen-ODE}) for $G_\lambda (x)$, and using this to find the generating function $F_\lambda (x)$, we get
\begin{equation}
\label{gen-func}
F_\lambda (x) = G_\lambda (0) x (1 + x)^{\lambda - 1} (1 - x)^{-\lambda - 1}.
\end{equation}
The constant $G_\lambda (0)$ acts as a scaling factor for the function, and therefore the associated sequence. Thus, the solution we obtain for the recursion relation is essentially unique for a given value of $\lambda$. Pre-classicality requires that there is no poles at $x=-1$; looking at the Taylor series coefficients of, e. g., $(1+x)^{-1} = 1 - x + x^2 + \cdots$ shows that it has undesirable oscillations. This results in the condition $\lambda > 0$. Once we have our generating function solution (\ref{gen-func}), we can use the binomial theorem for each of the factors, and write out its series expansion. Then we read off the coefficients of the powers $x^m$ to get the solution to the recursion relation:
\[
a_{\lambda > 0} (m) = G_\lambda (0) \sum_{j=0} ^{m-1} (-1)^j \binom{\lambda-1}{m-j-1} \binom{-\lambda - 1}{j}.
\]
This can be re-written in terms of hypergeometric functions as
\begin{widetext}
\begin{equation}
\label{a-def}
a_{\lambda > 0} (m) = G_\lambda (0) \biggl[ \binom{\lambda-1}{m-1} \HGeo (1-m, \lambda+1; \lambda-m+1; -1) - (-1)^m \binom{\lambda - 1}{-1} \binom{-\lambda-1}{m} {}_3\!F_2 (1, 1, \lambda + m + 1; m+1, \lambda + 1; -1) \biggr].
\end{equation}
\end{widetext}

\begin{figure}[hbt]
	\includegraphics[width=0.4\textwidth]{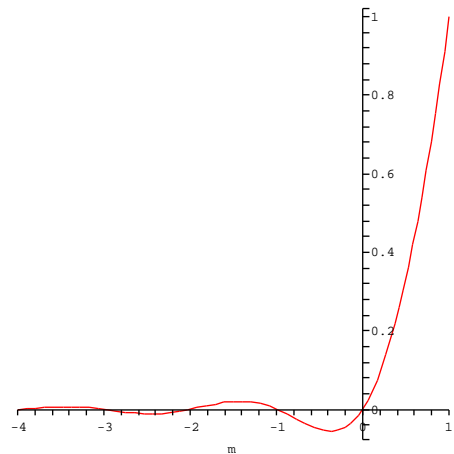}
	\caption{\label{fig1}Graph of the exact solution $a_{\lambda} (m)$ given in terms of hypergeometric functions by (\ref{a-def}) for $\lambda = 3/2$. For positive $m$, the sequence increases without limit. For negative $m$, there is an oscillatory decay in the wave function, representing a region where there is a small probability of finding the Universe, compared to the positive side of the classical singularity.}
\end{figure}

Even though this is coming originally from a Taylor series with positive integer values of $m$, we can use the formula (\ref{a-def}) as the {\it definition} of the sequence for {\it all} values of $m$, including the non-integers. An example of this exact solution is given in Figure \ref{fig1} for the value $\lambda = 3/2$. The sequence smoothly passes through the classical singularity $m=0$, and its behavior noticeably changes as it does so. Moving away from the singularity towards large positive $m$, the solution increases without limit; traveling from the singularity towards large negative $m$, the sequence undergoes oscillatory decay. Because these are the coefficients of the wave function describing the model, we can see that it is much more likely to find the Universe in the positive $m$ regime.

However, the recursion relation (\ref{a-eqn}) is only an approximation to the usual case of when the singularity is crossed, such as the "temporal" sequences $b_\lambda (n)$ for Bianchi I LRS, and $c_{\lambda_3} (m_3)$ for full Bianchi I. Examining the behavior of these sequences as they cross the singularity will give insight into the generic properties of the complete sequences $t(m, n)$ and $t(m_1, m_2, m_3)$, respectively, for these types of models. Although we now are studying separable sequences where the prime application is to the sequences $b_\lambda (n)$ and $c_{\lambda_3} (m_3)$, we will continue to use the notation $a_\lambda (m)$ as a generic name for all the solutions we find. 

\begin{figure}[hbt]
	\includegraphics[width=0.4\textwidth]{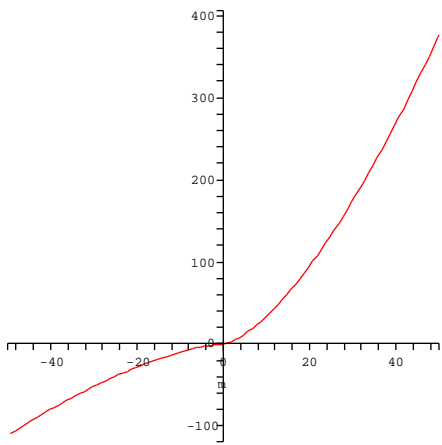}
		\caption{\label{fig2}Graph of the numerically evolved solution $a_\lambda (m)$ for $\lambda = 3/2$ and using the recursion relation (\ref{b-eqn}) with the function $d(m)$ and discrete parameters $m = k + 3/4$, $k$ an integer. Initial data is chosen for large positive $m$ using the exact solution (\ref{a-def}), and evolved using the recursion relation in the negative direction. Note the large difference in slope between the two sides of the graph for both sequences.}
\end{figure}

As mentioned previously, the non-polynomial nature of the function $d(n)$ appearing in the recursion relation is difficult to deal with using generating function techniques. We start with the exact values obtained by the definition (\ref{a-def}) for large (non-integer) positive values of $m$ and evolve numerically using equation (\ref{c-eqn}) with $d(m)$; for the results shown in Figure \ref{fig2}, we have chosen the specific parameters $\lambda = 3/2$ and $m = k + 3/4$, where $k$ is any integer. Notice that, as the sequence passes the origin, the slope of the line changes, but there is no decay. Due to the invariance of the difference equation as $m \to -m$, the sequence $a(k + \nu)$, where $k$ is an integer and $\nu \in [0, 1)$, is proportional to $a(k + (1 - \nu))$, but the ratio of the two varies with $\nu$. Because of this dependence, the sequences parametrized by $\nu$ should be thought of as independent, rather than part of a unified solution for all $m$. The sequences have a deflection because of the behavior of $d(m)$, differing greatly for $|m| < 1/2$ from a simple $1/m$ dependence. This does not happen for the integer $m$ sequences, since $d(0) = 0$, and the $|m| < 1/2$ region is avoided. These results, and the sequence seen in Figure \ref{fig2}, are similar to those present in discrete quantum gravity~\cite{gam-pul03}. Because we are seeing the same type of behavior for very different discretized version of loop quantum gravity, it is easy to suppose that these results are generic. Even though there may be modifications to the process of quantization -- such as the use of a self-adjoint constraint in group averaging techniques -- we can expect to see the same flavor of states as found here.

If we are focusing on one-dimensional sequences where the parameter is strictly non-integer, then we have not exhausted all the possibilities. We can consider the case when $a_\lambda (0) \ne 0$, which unfortunately makes the equation (\ref{gen-ODE}) for the generating function more complicated. There are two cases where sequences of this nature may arise. First, recall in the Bianchi I LRS model, when we simplify the Hamiltonian constraint under the definition $t (m, n) = \delta_m {\tilde t} (m, n)$, there is no restriction on the values of $t(0, n)$; however, because of the gauge conditions $t(-m, n) = t(m, -n)$ there is no need to look at the solution for negative values of $m$. Another possibility is recent work~\cite{nou-per-van05} using group averaging, where it was found that the physical Hilbert space consisted of sequences which were non-zero only at discrete values of the triad parameter $m$. Thus the sequence is typically zero at the classical singularity, yet if we find a pre-classical solution for all values, we can restrict this to only certain values and avoid the singularity in this manner.

\begin{figure}[hbt]
	\includegraphics[width = 0.4\textwidth]{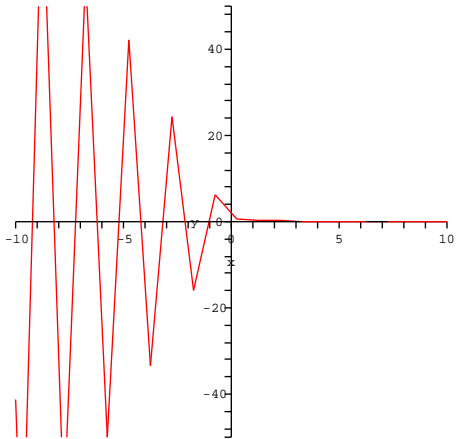}
	\caption{\label{fig3}Graph of the numerically evolved solution $a_\lambda (m)$ for $\lambda = -1$, $a(0) = 1$ and using the recursion relation (\ref{a-eqn}) with the function $d_2(m)$ and discrete parameters $m = k + 3/4$, $k$ an integer. Initial data is chosen for large positive $m$ using the exact solution (\ref{a-def2}), and evolved using the recursion relation in the negative direction. Unlike the previous graphs, the sequence becomes oscillatory with greatly increasing magnitude when $m < 0$. For positive $m$, the sequence is gradually decreasing, but much smaller magnitude (so it is not visible).}
\end{figure}

When $\lambda > 0$, pre-classicality gives no conditions on the sequence, and the results are similar to the $a_\lambda (0) = 0$ case -- the solutions grow without bound for increasing values of the parameter $m$. However, when $\lambda < 0$, there is a restriction relating the two values $a_\lambda (0)$ and $G_\lambda (0)$; when this is used, the associated generating function is given by~\cite{car-kha-boj04}
\[
\frac{F_\lambda (x)}{a_\lambda (0)} =   \biggl[1 + \frac{2^{\lambda} x (1-x)^{-\lambda-1} }{1 - \lambda} \HGeo(1-\lambda,-\lambda;2-\lambda;(1+x)/2) \biggr],
\]
Up to scaling by $a_\lambda (0)$, the sequence is fixed by the value of $\lambda$. As a simple case, we consider $\lambda = -1$, which gets rid of the $(1-x)^{-\lambda - 1}$ factor in $F_\lambda (x)$, with accompanying simplification in the Taylor series. Following the same procedure as with the $a(0) = 0$ situation, we get
\begin{equation}
\label{a-def2}
\frac{a_{-1} (m)}{a_{-1} (0)} = (-1)^{m} \biggl[(1 - 2m \ln 2) - 2 \sum_{j=0} ^{m-2} \frac{(-1)^j (m - j - 1)}{j+1} \biggr].
\end{equation}
This can be written as a complicated formula in terms of hypergeometric functions for all $m$, which we will not give here. However, this function is not defined at negative integers, so we only use it for positive $m$, then numerically evolve for negative $m$ with the recursion relation (\ref{a-eqn}). Because of the $(-1)^m$ factor, the function will be complex for non-integer $m$, but the ratio $a_{-1} (m+1) / a_{-1} (m)$ is real, so we can use the modulus of the function when finding the sequence values for positive $m$. The sequence gives behavior quite different from the $a_\lambda (0) = 0$ situation as the singularity is passed. This can be see in Figure \ref{fig3}, where the oscillations on the left hand side are huge compared to the smooth evolution of the curve on the right hand side (too small to be seen in the graph). This result is generic for $a_\lambda (0) \ne 0$ and $\lambda < 0$; sequences that are pre-classical for positive values of $m$ display large oscillations as the parameter becomes negative. The sequences are unable to match the symmetry of the recursion relation, and therefore are only smooth on one side of the singularity.

To summarize, we have examined separable solutions for the difference equations that appear as Hamiltonian constraints in LQC, and shown how to obtain them for all triad parameters, not just integer values. The large-scale behavior of these sequences is categorized by the separation constant $\lambda$ and value $a_\lambda (0)$ at the singularity. In particular, for $\lambda < 0$, the possible sequences displaying pre-classicality are very limited. When $a_\lambda (0) \ne 0$, the sequences are smooth only on one side of the singularity; they do not exist at all for $a_\lambda (0) = 0$. On the other hand, for positive $\lambda$, pre-classical sequences abound. There exists a one-parameter family of solutions when $a_\lambda (0) = 0$, and for $a_\lambda (0) \ne 0$, the sequences are defined by their values of $a_\lambda (0)$ and $G_\lambda (0)$.

These results have implications for the relationship between pre-classicality and the large volume limit. As discussed using solutions of the Wheeler-de Witt equation~\cite{car-kha-boj04}, to assemble an arbitrary semi-classical profile for the wave function at large volumes, one needs a complete basis of functions made up of all possible values of $\lambda$. For example, in Bianchi I LRS, when the temporal sequence $b_\lambda (n)$ is restricted to integer sequences, then pre-classicality requires $\lambda < 0$ (because of the relative sign between the recursion relations). This gives spatial sequences $a_\lambda (m)$ that decay to zero with increasing $m$. Non-integer sequences $b_\lambda (n)$, which do not pass through the origin and hence are not fixed to zero there, can get around this limitation. This allows a total sequence made of sums across all $\lambda$, so that any large volume behavior is feasible. The flip side is that this pre-classicality occurs only on one side of the singularity when $\lambda > 0$ (again, because of the sign flip for $b_\lambda (n)$).

Whether solutions to cosmological models should be pre-classical on both sides of the classical singularity or not is open for discussion. If one feels that in the distant ``past'' before the Big Bang, physics at scales comparable to the current scale of the Universe would necessarily be describable by partial differential equations, then pre-classicality on both sides of the singularity would be a physical requirement. In that case, the homogeneous models we have discussed here appear to lack the variety one would desire its pre-classical solutions to have. This would reinforce known results from the integer sector of the theory \cite{car-kha05, car-kha-boj04}. On the other hand, if one insists on pre-classicality only on one side of the singularity, then there is no such issue. Both cases results in an internal time asymmetry crossing the classical singularity -- either the scale of the wave function changes (so physical observables such as volume vary with time at different rates), or the Universe descends into a quantum foam with no classical behavior at all.

\acknowledgments

The authors appreciate the helpful comments of Martin Bojowald. GK is grateful for research support from the University of Massachusetts at Dartmouth, as well as Glaser Trust.

\end{document}